# Promoting the Research of Health Behavior Change in Chinese HCI Community

**Yunlong Wang**
Human-Computer Interaction Group
University of Konstanz
Konstanz, Germany
yunlong.wang@uni-konstanz.de

**Harald Reiterer**
Human-Computer Interaction Group
University of Konstanz
Konstanz, Germany
harald.reiterer@uni-konstanz.de

**ABSTRACT**

Unhealthy lifestyles largely contribute to many chronic diseases, which makes the research on health behavior change crucial for both individuals and the whole society. As an interdisciplinary research field, health behavior change research in the HCI community is still in the early stage. This research field is notably less developed in Chinese HCI community. In this position paper, we will first illustrate the research of health behavior change in the HCI community based on our previous systematic review. According to the unique properties of Chinese society, we will then discuss both the potential advantages and challenges of conducting health behavior change research in China. Lastly, we will briefly introduce the SMARTACT project in Germany to provide a reference for future related research. This paper aims to draw more attention to this research field and promote its development in China.

**KEYWORDS**

Health behavior change; Chinese HCI community

# 1 INTRODUCTION

Many lifestyle-caused chronic diseases - e.g., cardiovascular disease, type 2 diabetes, obese, and musculoskeletal pain - have led to heavy burdens on patients, their family, and the whole society. Accordingly, a new frontier of health behavior change research - precision lifestyle medicine [10] - is emerging to tackle the increasing burden of chronic diseases. The core of precision lifestyle medicine is health behavior change aiming to change unhealthy lifestyle behaviors before the onset of diseases. The research of health behavior change (HBC) refers to any theoretical and practical work on improving individuals' health-related behaviors to prevent diseases and improve health. Human behavior is complex, which involves habit, automatic responses to the immediate and wider environments, conscious choice and calculation, and is located in complex social environments and cultures [22]. Because of both its potential and complexity [22], academia and industry have invested many efforts into this field.

In the past decade, HBC increasingly attracted HCI researchers to this filed. From self-monitoring [1] to goal-setting [3], from action planning [18] to subliminal priming [14], from theory-based design [17] to glanceable user interface [2,5], HCI researchers contribute great work to the HBC research. Regarding industry, the most common products of HBC are fitness trackers and apps (e.g., Mi Band or Fitbit). Nevertheless, as an interdisciplinary field, the research of HBC is still in its infancy. In this paper, we propose to bring the HBC research to the conversation. We hope to promote the study and practice of HBC in Chinese HCI community.

# 2 HEALTH BEHAVIOR CHANGE IN HCI

The research of HBC is rooted in psychological theories (e.g., the Health Action Process Approach [16]) and public health practice (e.g., the Behavior Change Techniques [11]). Its inter-disciplinary nature brings both opportunities and challenges, which requires the collaboration of researchers from diverse backgrounds.

Along with the prevalence of electronic devices - especially personal computers and smartphones - researchers realized the potential of these devices for influencing our health in the long term. HCI researchers bring new methods of data collection, in-situ intervention, and user interaction analysis to the HBC research field. However, this field is still young and faced with many challenges. As HCI researchers, we used to focus on designing and evaluating new devices and software following the guidance of persuasive technology [4,12]. Meanwhile, we are also likely to ignore the behavior change theories [7,19], which is of fundamental importance in behavior change research. It is similar in the counterparts of public health and psychology, where the lack of user experience consideration is typical (e.g., see [20]). As an initial attempt, we proposed a holistic framework for HBC researchers to design, evaluate, and report their related study in a formative and comprehensive way [17].

To reveal the status of the HBC research in the HCI community, we conducted a systematic review of the related work over the 20-year development [19]. Based on the ACM digital library, we found 648 related papers. After a 10-year increase, the single year's paper amount peaked at 128 in 2016. We screened the papers and obtained 75 empirical studies for in-depth review. Our main findings of trends and patterns of the HBC research in the HCI community include:

- The target behaviors mainly fell into the physical activity category, which was followed by dietary behavior, sleeping, and smoking. Some critical behaviors (e.g., sedentary behavior [20] and stress management [15]) were much less addressed.
- The target users or study participants were mostly adults who were mostly college students and staff. We believe that more attention on the elderly is required as many countries have an increasingly aging population.
- There is no standardized approaches or instruments for evaluating the user experience of intervention systems for HBC.
- There is no standard to report the intervention study for HBC, which hinders in-depth systematic reviews for knowledge accumulation.

Based on the review, we also see recent emerging research themes such as gamification, social network, and virtual agents in this field. We believe the HBC research in the HCI community will keep growing due to the high demand and the many unsolved research questions.

## 3   HEALTH BEHAVIOR CHANGE AND MOBILE HEALTH IN CHINA

As far as we know, the research of HBC in Chinese HCI community is minimal. Among the 75 selected studies in our review [19], no one was conducted in mainland China. Although the article database limits our review, the result is in line with another systematic review [13], reflecting the lack of HBC research in the HCI community in China.

It should be noted that HBC is not identical to digital health (eHealth) or mobile health (mHealth), which is a broader concept referring to any medical and public health practice supported by digital devices or mobile devices, respectively [21]. Although HBC research is very limited in China, mobile health applications and services keep growing in the Chinese market.

Although this paper is not focused on mHealth applications, it is worthy to highlight two related reviews to get some ideas about the overall mHealth development in China. In 2014, Li and colleagues reviewed the mHealth development in China, especially the mobile health applications in treating mental illnesses and teleconsultation [9]. Their review showed that the mHealth services were in the initial stage but the demand is high. Besides, they shed some light on the challenges of mHealth development in China - e.g., national regulation and standards of mobile health services were lacking. In 2015, Hsu and colleagues selected and reviewed 234 Chinese mobile health apps and found that the most common mHealth initiatives were telemedicine, appointment making, and medical education [8]. Regarding the target disease or behavior, diabetes, hypertension, and hepatitis were the most common ones these apps were designed for. The review of Hsu et al. also mentioned the lack of corresponding regulation and standards of mobile health services.

To sum up, the Chinese market has a high demand for mHealth services. The mHealth applications (including HBC) developed at a higher pace in industry than in academia.

## 4   PROMOTING HEALTH BEHAVIOR CHANGE RESEARCH IN CHINA

Despite the current status of the HBC research in the HCI community in China, we believe it holds great potential due to the health culture, the technology prevalence, and the demanding market in China.

Chinese people are culturally prepared for health promotion. Keeping in good health (养生 in Chinese), a concept rooted in traditional Chinese medicine, has been part of Chinese culture over two thousand years. Chinese people traditionally follow some health-guidelines in their lives - e.g., eating different food according to the season and their health states. Another example is keeping the balance of physical activity - they believe that too much viewing, lying, sitting, standing, or walking could injure our bodies in different ways. This theory was explained in a classic of traditional Chinese medicine - Esoteric Scripture of the Yellow Emperor (皇帝内径 in Chinese). Therefore, with the health culture, it could be convenient to motivate people to participant in health promotion studies in China.

The technology acceptance rate and speed are high in the Chinese market. New technologies might help us to improve health problems; it might also bring us extra health problems. For example, the high participation rate of the social network can enable the research of social impact on physical activity [6]. In parallel, the social network could also cause prolonged online gaming time among teenagers. Therefore, there could be more research questions and opportunities for HBC researchers in China. Moreover, the high integration of mobile services (e.g., food ordering, traveling, physical activity, entertainment) also provides the opportunity to study HBC in a holistic view.

The vast user groups in the Chinese market have diverse needs for HBC research and products. For example, parents need new technology to reduce their children's screen-based activity. Office workers need new technology to promote their mental and physical health under high working pressure. Aging people need new technology to manage health conditions and obtain healthcare services.

Besides the market opportunities, HCI researchers in China have advantages regarding cross-discipline collaboration. Chinese universities always have diverse subjects including natural science, social science, medicine, and engineer, which objectively encourages inter-disciplinary research. The high density of universities and institutes in mega-cities in China (e.g., Beijing or Shanghai) could ease the cross-university collaboration. Moreover, many universities have affiliated hospitals, which is another convenience for HBC research.

## 5  CONCLUSION

The spirit of the present paper is driven by the authors' passion for HBC research, which might benefit thousands of people's life quality. We hope the discussion could rise the interests and actions of Chinese HCI community.

## ACKNOWLEDGMENTS

The authors are involved in the SMARTACT project (https://www.uni-konstanz.de/smartact/), which is an ongoing collaborative research project (2015-2021) aiming to improve the long-term health behavior by using mobile technologies (e.g., smartphones). It is funded by the German Federal Ministry of Education and Research (BMBF) and consists of research groups of HCI, psychology, and economics from three universities in Germany. Our goal in SMARTACT project is to develop and evaluate a module-based toolbox to study and improve users' eating behavior and physical activity through the lenses of individual factors (e.g., motives, emotions, stress, goal-setting, and action planning) and social context triggers (in the family and the job context). The scientific advisory board includes 13 international professors in the domains of health science, computer science, psychology, economics, and public health. We would like to introduce this project in the workshop.